\documentclass [reviewcopy]{elsart}

\usepackage{graphicx}

\usepackage{amssymb}

\begin{document}
\begin{frontmatter}


\title{Spin dynamics in  exchange-biased F/AF bilayers
}
\author{D. Spenato\corauthref{cor}}
\corauth[cor]{Corresponding author: LMB/UBO/CNRS UMR 6135, 6
avenue Le Gorgeu, 29285 Brest, France - fax +33 298 017 395 }
\ead{david.spenato@univ-brest.fr}
\author{J. Ben Youssef and H. Le Gall }
\address{Laboratoire de Magn\'etisme de Bretagne, UBO/CNRS/UMR 6135, 6 avenue Le Gorgeu, 29285 Brest,
France}




\begin{abstract}
The spin dynamics of the ferromagnetic pinned layer of
ferro-antiferromagnetic coupled  NiFe/MnNi bilayers is
investigated in a broad frequency range (30 MHz-6 GHz). A
phenomenological model based on the Landau-Lifshitz equation for
the complex permeability of the F/AF bilayer is proposed. The
experimental results are compared to theoretical predictions.
\end{abstract}

\end{frontmatter}

\newpage

\section{Introduction}

Exchange coupling  between a ferromagnetic (F) and an
antiferromagnetic (AF) layer has been extensively studied in last
few years because of its importance for pinning the ferromagnetic
layer in giant magnetoresistance spin-valve. MnNi seems to be a
promising alloy for exchange coupling to NiFe due to its large
exchange coupling field, its good corrosion resistance and its
high blocking temperature \cite{Lin}. Irreversible measurements
(such as hysteresis loop) are commonly used to quantify the
unidirectional exchange coupling in F/AF bilayers. Reversible
experimental techniques, such as ac susceptiblity and
ferromagnetic resonance (FMR) and Brillouin light scattering
(BLS) are also employed (for a review see \cite{revex}).  These
techniques lead to different values for the exchange field
\cite{fermin}. Most of the reversible techniques involve a
constant frequency excitation and only a few papers \cite{byeon}
deal with the measurement of the complex susceptibility of F/AF
bilayers in a broad frequency range. The aim of this paper is to
probe, with the measurement of the complex permeabiliy frequency
spectrum, the exchange anisotropy of NiFe/MnNi bilayers with
different Mn concentrations.

\section{Experiments}

Substrate$\backslash$$Ni_{81}Fe_{19} 280$\AA
$\backslash$$Mn_{x}Ni_{100-x} 800$$\AA$  bilayers were grown on
Corning Glass substrate  by RF diode sputtering using a standard
Z 550 Leybold equipment with a magnetic field of 300 Oe applied
during deposition to induce an uniaxial anisotropy. The
background pressure was lower than $4\times10^{-7}$ mbar. Ni chips
were homogeneously added to a four inches diameter Mn target in
order to get films in the Mn composition range 35-80 percent. The
chemical homogeneity was verified by Electron Probe Micro
Analysis (EPMA) on several points of the sample. The Mn
composition variation is about one percent on the entire sample.
After deposition, samples were annealed in a magnetic field of
1000 Oe, aligned with the easy axis of the film, at $300^o$C for
5 hours to induce the exchange field. The magnetic properties
such as the saturation magnetization  $M_{s}$ were obtained from
magnetization loops (M-H loops) measured at room temperature
using a VSM. The complex frequency spectra of the bilayers were
measured from 30 MHz to 6 GHz using a broad band method. The
method is based on the measurement, by a network analyser, of the
reflection coefficient $S_{11}$ of a single turn coil  loaded by
the film under test \cite{monos1}. Because of the topography of
the applied ac field ($h_{ac}$) in the coil, the permeability can
be measured for different orientations of the exciting field in
relation to the in-plane anisotropy.

\section{Results and discussion}

We consider a F layer, with an uniaxial anisotropy field H$_{k}$
along the easy axis, submitted to an unidirectionnal exchange
field H$_{ex}$, induced by the exchange coupling along the easy
axis, and to an applied ac field $h_{ac}$ applied perpendicular
to the easy axis. The initial susceptibility measured along the
ac field is given by \cite{xi} $\chi=M_{S}/(H_{ex}+H_{k})$.
$\chi/M_{S}$ is the initial slope of the hysteresis loop when
measured perpendicular to the easy axis. $H_{ex}$ is determined
by the shift of the center of magnetization loop. $H_{k}$ is
extracted from the M-H loops measured perpendicular to the easy
axis (previous equation).

In $Ni_{81}Fe_{19} /Mn_{x}Ni_{100-x}$ bilayers the exchange field
depends strongly on the Mn concentration  and the growth
conditions \cite{dave1}. We have grown several bilayers, whose
composition varies from 35 to 80 percent of Mn. Before annealing,
for all the samples, only the soft magnetic behavior of NiFe was
detected. After annealing and above 40 percent of Mn, the MnNi has
gone though a Paramagnetic-AF phase transition. The M-H loops are
shifted and typical of a F/AF coupling. The NiFe layer is still
showing an in-plane uniaxial anisotropy. We have measured
$H_{k}^{mes}$ and $H_{ex}^{mes}$ on samples with different Mn
compositions. The results are presented in Table \ref{table}. One
can observe that the saturation magnetization of the NiFe layer
decreased after annealing. This may be attributed to interfacial
diffusion leading to the existence of an interdiffused AF ternary
alloy FeMnNi at the NiFe-MnNi interface \cite{dave1}. It can be
seen that the values of the exchange field increase as a function
of the Mn composition. The values of the anisotropy field
$H_{k}^{mes}$ (up to 3 kA/m) are larger than those of the
uncoupled annealed NiFe layer, and these values change
significantly with the Mn composition.

We have measured the complex permeability spectra of the bilayers
before and after annealing  for the exciting field applied
perpendicular to the easy axis. The results are presented in
fig.\ref{figure1} for three samples and the major results for all
the samples are presented in fig.\ref{figure2} . Before annealing
(as grown), there is no magnetic interaction between the NiFe and
the MnNi layer. The permeability spectra are typical of damping by
spin rotation processes in a NiFe layer \cite{oliv}. For the
annealed samples, when the exchange and the anisotropy fields
increase, the level of the real part of the permeability
$\mu'(0)$ at low frequency decreases and the roll off frequency
increases. We can also observe that the imaginary part of the
complex permeability $\mu''$ shows a lower resonance peak, a
higher resonance frequency $f_{res}$(up to 3.5 GHz) and a wider
resonance peak as the exchange field increases.

In a previous paper \cite{dave2}, we have presented an analytic
calculation of the frequency dependent complex permeability
tensor of a thin ferromagnetic film with uniaxial in-plane
anisotropy, submitted to an external exciting field using the
Landau-Lifshitz (LL) theory \cite{landau}. Using this calculation,
we have obtained the components of the complex permeability tensor
which are a function of $M_{s}$, the total effective field
$H_{eff}$, the frequency f of the exciting field and the
phenomenological damping constant $\alpha$. The theoretical value
of $\mu'(0)$ at low frequency is found to be 1+($M_{s}/H_{eff}$)
and, as observed, the decrease of the saturation magnetization
and the enhancement of the effective field
($H_{k}^{mes}+H_{ex}^{mes}$) lead to a reduction of the level of
$\mu'(0)$. The resonance frequency $f_{res}$ is found to be
(1/2$\pi$)$\times$$\gamma$($H_{eff}(H_{eff}+M_{s}))^{1/2}$. In
our samples the enhancement of $H_{eff}^{mes}$ is prevalent and
lead to the enhancement of $f_{res}$. Fig. \ref{figure3} shows an
example of the comparison between theoretical and experimental
complex permeability spectra when the exciting field is applied
perpendicular to the easy axis. The results are presented for an
as-grown sample and for one exchange-biased sample
(NiFe$/Mn_{46}$Ni$_{54}$). For the other samples the fit
parameters are presented in Table \ref{table}.

In a first step of calculations, the values the effective field
($H_{k}^{mes}+H_{ex}^{mes}$) and $M_{s}^{mes}$  are taken from
static measurements and the value of the damping parameter is
fitted. For the as-grown sample $(H_{ex}^{mes}=0)$, experimental
results are in good agreement with the theoretical prediction as
observed in fig. \ref{figure3} (solid curve (a)). The value of the
fitted damping parameter (0.012) is typical of the one obtained
on a NiFe single layer \cite{oliv}. For the exchange biased
bilayers, the level of the permeability is in agreement with the
theoretical values but the calculated resonance frequency is
lower than the measured one (Fig.\ref{figure3} solid curve (a)).
In a second step we have computed the complex permeability where
the effective field was taken as a fit parameter
($H_{eff}^{fit}$). The result is presented in figure \ref{figure3}
(solid curve (b)). It can be seen that the experimental results
are in good agreement with theoretical predictions. The fitted
values of the effective field in the exchange biased bilayers are
higher that those obtained from M-H loop measurements. Moreover,
one can see that the broadening of the experimental spectrum of
$\mu''$ is associated with the increasing of the damping
parameter from 0.012 up to 0.035 (Table \ref{table}). It may be
explained as follows. The experimental data should be not
interpreted with a phenomenological model of fixed moment along
the easy anisotropy direction. One should take into account a
local variation of the exchange field at the F/AF interface. This
local variation may be due to interdiffusion at the NiFe-MnNi
interface. This complex magnetic structure at the interface may
be grains of NiFe exchange coupled with antiferromagnetic NiFeMn
"coating" on their surface \cite{diff}. These broadenings have
been observed with FMR and BLS measurements and attributed to a
relaxation mechanism based on two-magnon scattering processes due
to the local fluctuation of the exchange coupling caused by
interface roughness \cite{rezende1}. In conclusion, we have shown
that it is possible to describe the magnetization dynamics of
exchange biased bilayers with the LL theory. Effective fields are
extracted from complex permeability spectra and are much larger
than the ones obtained from a hysteresis loop measurements. The
high values of the effective field associated with the enhancement
of the damping parameter may be associated with interdiffusion in
the bilayers.

This work was partially supported by PRIR program of Region
Bretagne

 \clearpage

\bibliographystyle{elsart-num}
\bibliography{biblio}

 \clearpage

 \begin{table}
\caption{Magnetic properties of the $NiFe/Mn_{x}Ni_{100-x}$
bilayers. Magnetization and field units are in (A/m)}
\label{table}
\begin{tabular}[h]{ccccccc }
x & $M_{s}^{mes}$ & $H_{k}^{mes}$&$H_{ex}^{mes}$& $H_{eff}^{mes}$&
$H_{eff}^{fit}$&$\alpha$\\
\hline

as grown &800$\times10^{3}$&360&0&360&360&0.012\\
46&711$\times10^{3}$&330&268&598&1010&0.027\\
58&662$\times10^{3}$&975&624&1599&3600&0.03\\
68&537$\times10^{3}$&3194&2082&5276&8500&0.035

\end{tabular}
\end{table}

\clearpage

\begin{center}
\section*{Figure Captions}
\end{center}

\begin{figure}[h]
\begin{center}
\caption{imaginary (a) and real part (b) of the complex
permeability spectra of  As grown  and annealed NiFe/MnNi bilayers
with different values of the exchange field} \label{figure1}
\end{center}
\end{figure}

\begin{figure}[h]
\begin{center}
\caption{Experimental resonance frequency ($\bullet$) and inital
permeability $\mu'(0)$ ($\circ$) as a function of the Mn
composition in NiFe/MnNi bilayers} \label{figure2}
\end{center}
\end{figure}

\begin{figure}[h]
\begin{center}
\caption{Measured and calculated $\mu''$ spectra of two NiFe/MnNi
bilayers when the exciting field is applied along the hard axis;
Solid line (a) simulated curve, (b) fitted curve} \label{figure3}
\end{center}
\end{figure}

\clearpage

Figure \ref{figure1} \hspace{2cm} D. Spenato 

\begin{figure}[h]
\begin{center}
\vspace{1cm}
\includegraphics[width=8 cm]{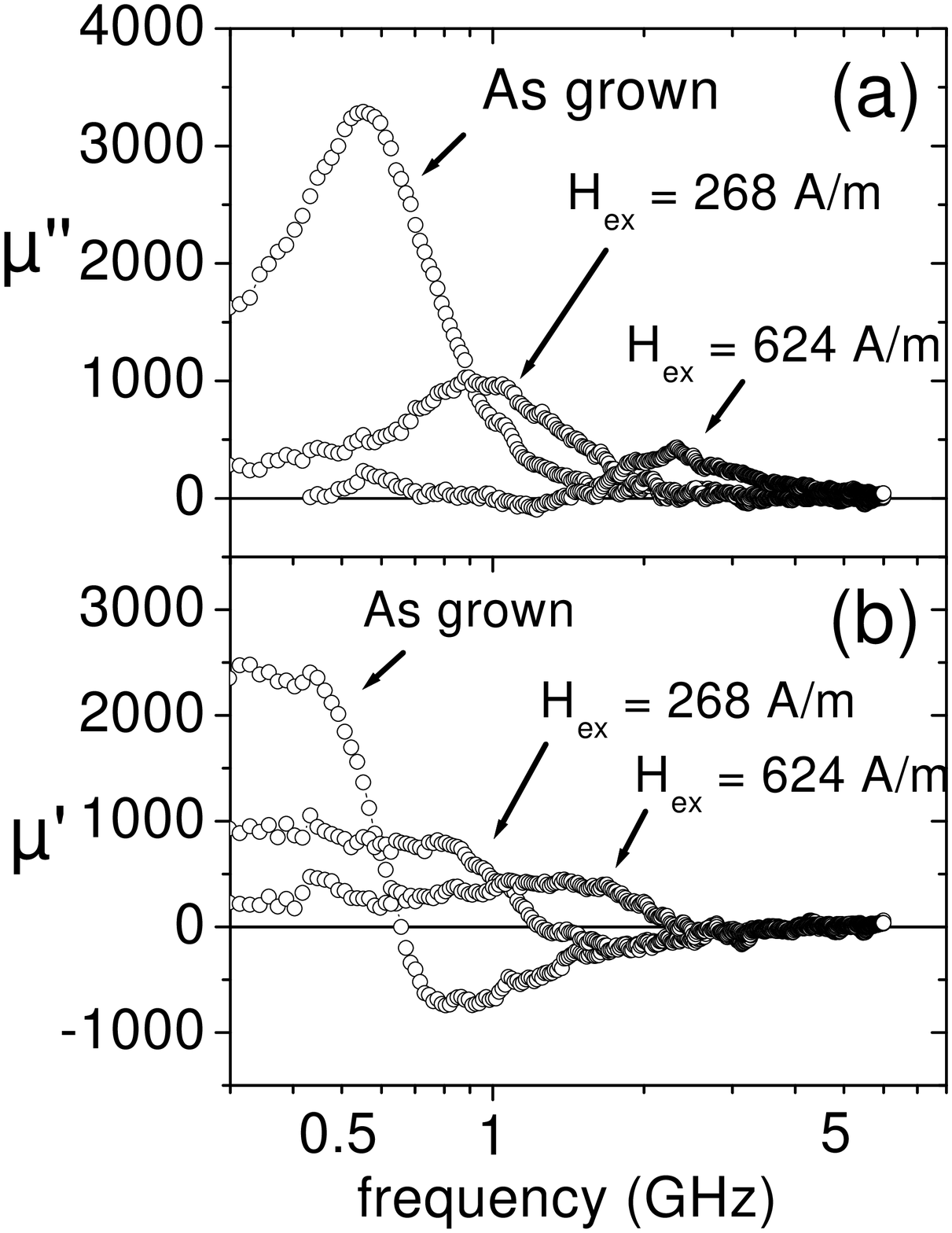}
\vspace{4cm} \\
\end{center}
\end{figure}

\clearpage

Figure \ref{figure2} \hspace{2cm} D. Spenato 

\begin{figure}[h]
\begin{center}
\vspace{3cm}
\includegraphics[ width=11 cm]{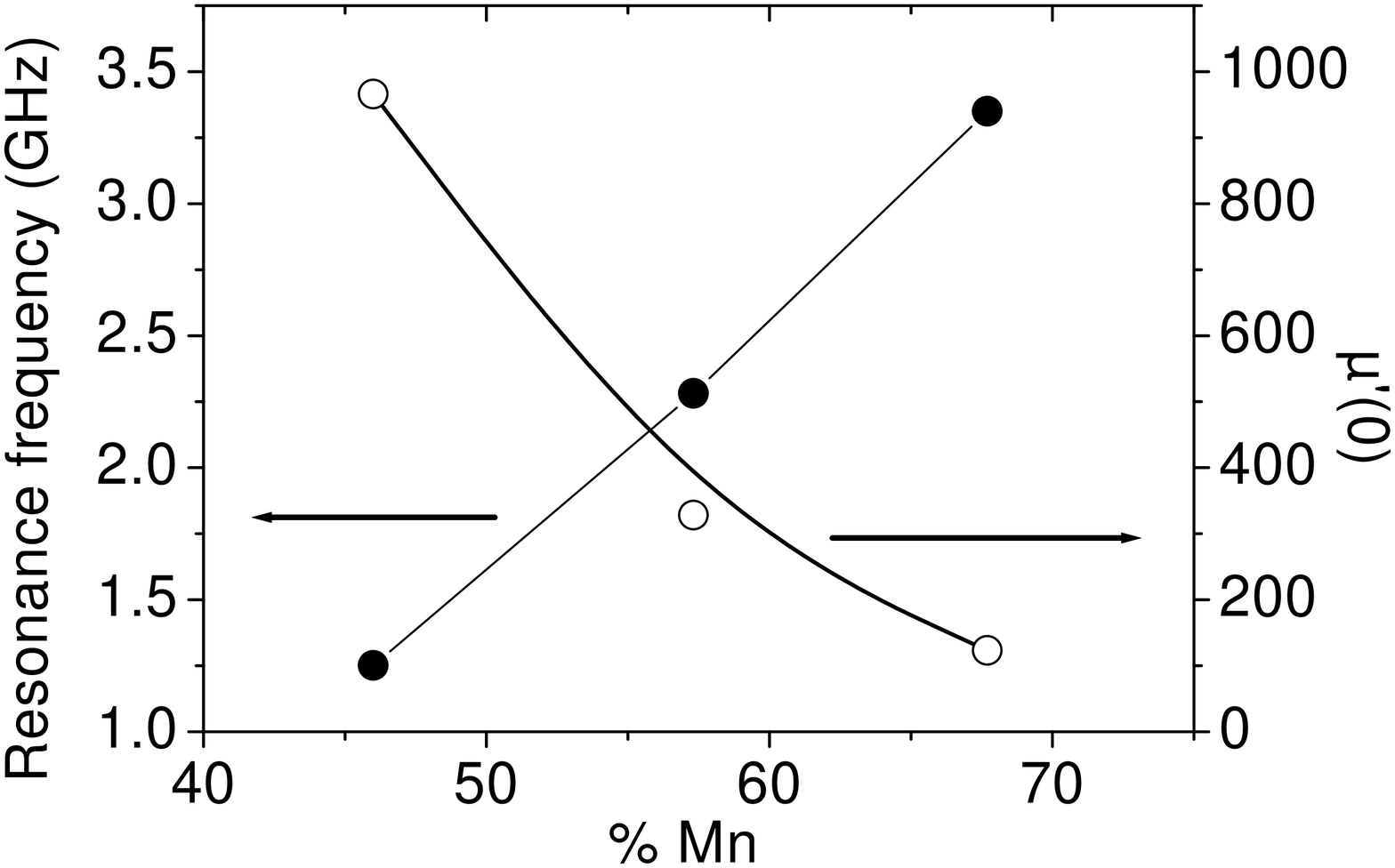}
\vspace{2cm} \\

\end{center}
\end{figure}

\clearpage

Figure \ref{figure3} \hspace{2cm} D. Spenato 

\begin{figure}[h]
\begin{center}
\vspace{3cm}
\includegraphics[ width=11 cm]{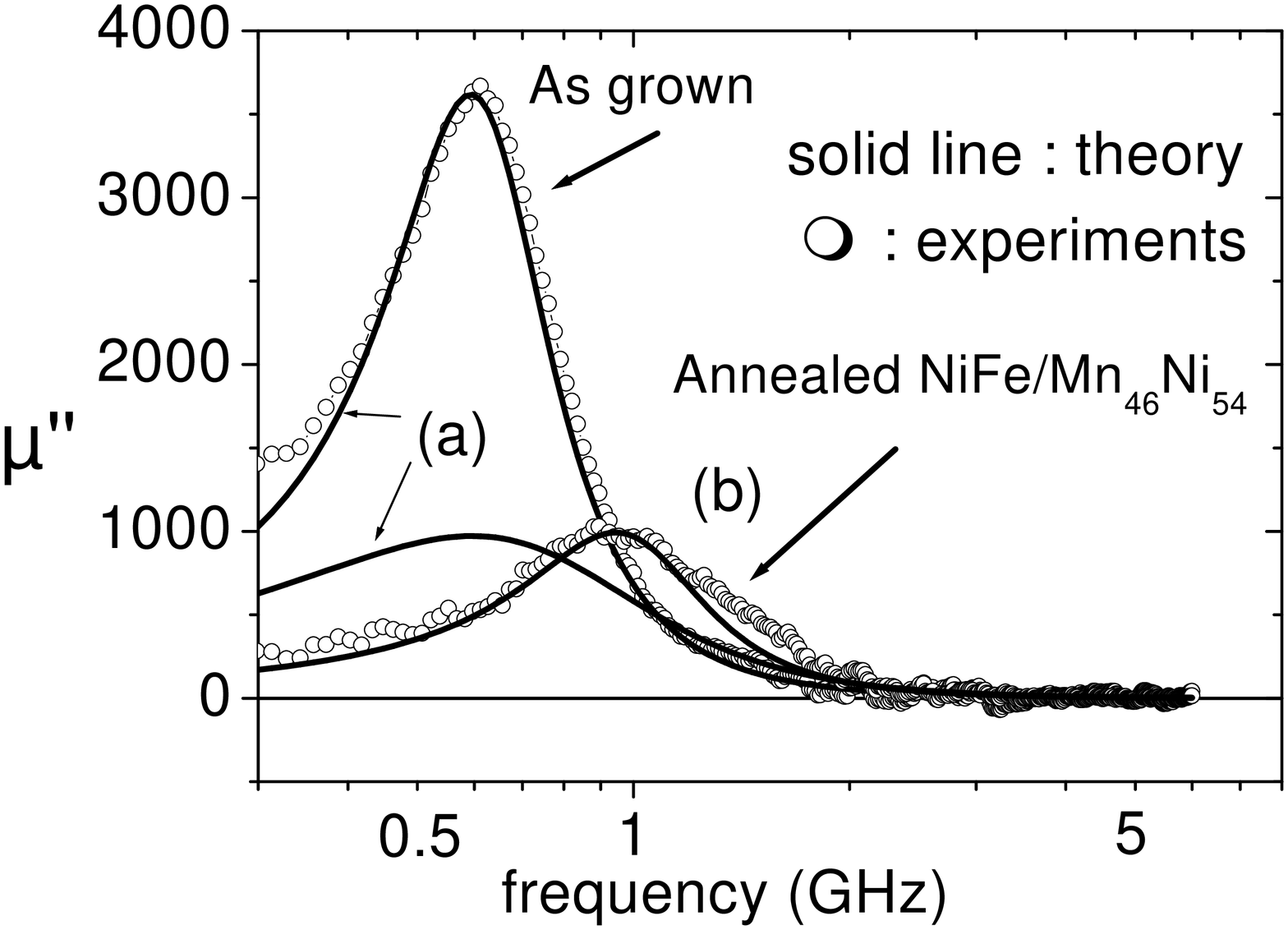}

\vspace{2cm} 
\end{center}
\end{figure}







\end{document}